\documentclass{article}%
\usepackage{amsmath}%
\usepackage{amsfonts}%
\usepackage{amssymb}%
\usepackage{graphicx}
\begin{document}

\title{The variational formulation of the theory of non-stationary propagation of a femtosecond laser radiation}
\author{Andrey D.Bulygin}
%\author{Andrey D.Bulygin $^{1,*}$}
%\\V.E. Zuev Institute of Atmospheric Optics SB RAS (IAO SB RAS), Tomsk, 634055, Russia}
%$^1$\vtop{\footnotesize  V.E. Zuev Institute of Atmospheric Optics SB RAS (IAO SB RAS), Tomsk, 634055, Russia}\\

%\date{February 24, 2017}
\maketitle

% \englishkeywords{nonlinear Schrodinger equation,variational formulation,virial theorem,conservation laws,filamentation,attractor}

\begin{abstract}
In this paper, an inverse variational problem is solved for the nonlocal nonlinear Schrödinger equation used in modeling filamentation in various nonlinear media.
The corresponding integral relations are found which generalize the conservation laws for the non-conservative case.
\end{abstract}
 \textbf{ key words}: nonlinear Schrodinger equation,variational formulation,virial theorem,conservation laws,filamentation 
%  \keywords{nonlinear Schrodinger equation,variational formulation,virial theorem,conservation laws,filamentation,attractor}

% \textbf{}
\section{Introduction}
The phenomenon of self-focusing and filamentation of high-power laser radiation (HFLR) is most often described on the basis of the so-called nonlinear Schrodinger equation (NSE) [\ref{Boyd}]. The variational formulation of this equation, in application to the problem of filamentation (with a specific type of equations for the medium) is widely known in the literature solely for stationary approximation. Within this stationary approximation, only a very limited set of rigorous analytical results is known [\ref{Boyd}]. This property of the global collapse in the Kerr medium, Townes' solution (which is also not analytic [\ref {taunes_profile}]) and the Bespalov-Talanov instability [\ref {Boyd}]. One way or another, at the moment there are actually not exist known rigorous properties of non-stationary (NES), which would allow, for example, to evaluate the correctness of the numerical solution.
In this paper we propose a procedure for constructing the Lagrangian formulation of these equations (and hence also the Hamiltonian formulation)\footnote{ The transition from the Lagrangian to the Hamiltonian formulation in the case of a nondegenerate theory is realized by means of the Legendre transformation, otherwise, i.e. in the theory with constraints, the transition is carried out on the basis of the Dirac-Bergman algorithm [\ref{DB}]}.  
The constructing the Hamiltonian formulation allows not only to automatically write out the integrals of motion for non stationary models,  but also provides a tool that allows the canonical implementation of the program of stochastic quantization in the application to the problem of filamentation. This is relevant because, generally speaking, for ultra-high-power laser radiation filamentation is a fundamentally stochastic process [\ref{Kand}].

\section{A nonstationary NSE with a nonlocal complex potential}
The non stationary model of filamentation takes into account the effects of dispersive spreading of the laser pulse,  the effect of plasma accumulation over time, cubic nonlinearity and higher-order nonlinearities [\ref{Boyd}].  In the so-called concomitant coordinate system, the equation for the complex envelope of the intensity of the light field is written in the Fourier representation in the form $[\ref{couiran}]$:

\begin{equation}
\label{eq: basic_eq_1}
T_u {\equiv  } i\frac {\partial }{\partial z }U+{(\partial_\mu\partial^\mu+\epsilon_{ker}[I]+\epsilon_{pl}[I]) U}+i \alpha_{I} U=0\\
\end{equation}

Here $\epsilon_{ker}[I]$ functional of the intensity, which we in accordance with the works $[\ref{couiran}]$ , we choose in the form:

\begin{equation*}
  \epsilon_{ker}(t)=n_{ker} (I+\epsilon_{in}[I])
\end{equation*}

where 
\begin{equation*}
\epsilon_{in}=\int_{-\infty}^\infty \theta(t-\tau) R (t-\tau)I(\tau) d\tau.
\end{equation*}

Here, the molecular response in the Kerr effect is modeled by the Green's function of a damped oscillator: $R(t)=exp(-\gamma_R t)sin(\Omega_R t)$.
Equation for $\epsilon_{pl}$ caused by the formation of a plasma under the action of laser radiation has the form: 

\begin{equation*}
 \frac {\partial \epsilon_{pl} }{\partial t }=({\epsilon_{pl}^0-\epsilon_{pl}})\psi_{pl}(I)+\psi_{pl}^1(I,\epsilon_{pl})
\end{equation*}

Here the ionization coefficient  $\psi_{pl}$ is a nonlinear function of the intensity and can be calculated, for example, according to the Popov-Perelomov-Terentyev model  $[\ref{ppt}])$;  $\epsilon_{pl}^0$ - the dimensionless concentration of molecules in the medium; $\psi_{pl}^1$ - The term by means of which takes into account such mechanisms as cascade ionization and recombination .

The set of equations $(\ref{eq: basic_eq_1})$ can be represented in the form
 \begin{equation}
\label{eq:nst_basic_eq}
 \left.\begin{aligned} 
T_u {\equiv  } i\frac {\partial }{\partial z }U+{(\partial_\mu\partial^\mu+\epsilon_{ker}[I]+\epsilon_{pl}[I]) U}+i \alpha_{I} U=0\\
T_{\epsilon_{pl} } {\equiv } \frac {\partial \epsilon_{pl} }{\partial t }-({\epsilon_{pl}^0-\epsilon_{pl}})\psi_{pl}(I)+\psi_{pl}^1(I,\epsilon_{pl})=0\\
T_{\epsilon_{in} } {\equiv } \frac {\partial^2 \epsilon_{in} }{\partial t^2 }+\gamma_R\frac {\partial \epsilon_{in} }{\partial t}+\Omega_R^2 \epsilon_{in}=I
 \end{aligned} \right\}
  \end{equation}

This system of equations $(\ref{eq:nst_basic_eq}) $, generally speaking, is not Lagrangian, since the necessary condition for it is not satisfied
 \begin{equation*}
\frac{\delta }{\delta \epsilon } T_u \neq \frac{\delta}{\delta u }  T_{\epsilon  }
\end{equation*}
 
Although, for the inertial part of the Kerr nonlinearity, the following condition is satisfied:

 \begin{equation*}
\frac{\delta }{\delta \epsilon_{in} } T_u = \frac{\delta}{\delta u }  T_{\epsilon_{in}  }
\end{equation*}
 
nevertheless the equation itself $T_{\epsilon_{in}}$ It is not Lagrangian, since this is an equation with friction, namely, an equation of the type of a damped oscillator.

Thus, the non-Lagrangianity of the system $(\ref{eq:nst_basic_eq}) $  is due to two circumstances. First, the equations for nonlinear dielectric permittivity contain the first time derivatives, this applies to both the inertial Kerr part and the plasma part. Secondly, the term in the equation for the light field, which causes interaction with the medium, does not have a corresponding partner in the equation for the medium, i.e. $\partial_{\epsilon}T^{int}_u \neq  \partial_{u}T^{int}_{\epsilon} $. 

The solution of the first problem is presented in a general form in the paper [\ref{Kup}]. Thus, considering the intensity of the laser field as the driving force in the right-hand side of the equation for the harmonic damped oscillator, to which the inertial Kerr nonlinearity is modeled, we can, respectively, write the action for the inertial term in the form: 
\begin{multline*}
S_{in}= n_{ker} \int e^{ \gamma_R t} ( (\dot{\epsilon_{in}} p_{\epsilon}-\epsilon_{in} \dot {p_{\epsilon}} -(p_{\epsilon}^2+\gamma_R  \epsilon_{in} p_{\epsilon}+\epsilon_{in}^2\Omega_R^2))/2 +\epsilon_{in} I(t)) d \vec{x} dz.
%S_{in}\,\BRK= n_{ker} \int e^{ \gamma_R t} (-\dot{\epsilon_{in}} p_{\epsilon}+\epsilon_{in} \dot {p_{\epsilon}} +(p_{\epsilon}^2+\gamma_R  \epsilon_{in} p_{\epsilon}+\epsilon_{in}^2\Omega_R^2))/2 \BRK+
%\epsilon_{in} I(t)d \vec{x} dz.
\end{multline*}
Where $p_{\epsilon}=\partial \epsilon/ \partial t$.

 We will notice that to receive the initial system of the equations it is necessary to increase together full Lagranzhian systems by this integrating multiplier 
(because of presence of the member of interaction) that corresponds to multiplication of all equations by this multiplier, with only that remark that to compensate 
the excess composed  arising in the equations for the field from members in the form of $ { \gamma_R t } \partial_t U \partial_t U^* of $ it is necessary to add to action the member in shape  $\gamma_R(\partial_t U  U^*-\partial_t U^*  U) e^{ \gamma_R t}  $,   what leading to actually to physically equivalent system of the equations. However given compensating composed is purely imaginary that represents some complexity for physical 

The situation with the equation for the plasma is somewhat different, to find such an integrating factor that would ensure the Lagrangianity of the equation for the plasma itself and compatibility with the equation for the light field, it seems to us possible, only by approximate perturbative methods. However, the practical value of this approach in this case is extremely small.
n this situation, it seems reasonable to introduce an auxiliary field. From physical considerations, one can guess which auxiliary field should be introduced in order to ensure the Lagrangianity of the complete system of equations. The remark can serve as a remark that the dynamics of electrons, in general, should be described by an equation for a complex field having the meaning of the probability density amplitude. Thus, it is natural to try to change from the real function of the plasma density to the complex value ${\chi}$, which has the meaning of the amplitude of the density of the number of particles participating in the energy exchange with the light field, which satisfies the Schrodinger equation and is related to the electron density $\epsilon_{pl} $, as     ${\chi \chi^*=\epsilon_{0}-\epsilon_{pl}}$.  Then it is easy to verify that the system of equations $(\ref{eq:nst_basic_eq}) $ is equivalent to the following system of equations: 
 
\begin{equation}
\label{eq:nst_basic_eq_2}
 \left.\begin{aligned} 
T^{(0)}_u +i \alpha_{I} U  {\equiv  } i\frac {\partial }{\partial z }U+{(\partial_\mu\partial^\mu+\epsilon_{ker}[I]+\epsilon_{pl}[I]) U}+i \alpha_{I} U=0\\
T^{(0)}_{\chi} +i \alpha_{\chi} \chi {\equiv } i\frac {\partial \chi}{\partial \tau_m }+\chi I+i \alpha_{\chi} \chi =0\\
T^{(0)}_{\epsilon_{in}}+\gamma_R\frac {\partial \epsilon_{in} }{\partial t} {\equiv } \frac {\partial^2 \epsilon_{in} }{\partial t^2 }+\gamma_R\frac {\partial \epsilon_{in} }{\partial t}+\Omega_R^2 \epsilon_{in}=I
 \end{aligned} \right\}
 \end{equation}
  
For a given system of equations, it is no longer difficult to write down an action: 
 
\begin{equation}
\label{eq:basic_S}
S_c=S_u^k+S_{\chi}+S_{in}+S_{kin}+i S_f+i S_{in}. 
\end{equation}

Here we have selected a part of the action, so that the variation of this part gives a derivative with respect to the evolution variable z for the field U and the nonlinearity of the Kerr type:
\begin{equation*}
 S_u^k=\int (i (\frac{\partial }{ \partial z}U ) U^* - i (\frac{\partial  }{ \partial z}U^*) U +\epsilon_k I^2/2)e^{ \gamma_R t}  d\vec{x}  dz.
\end{equation*}
The term in action responsible for the kinetic part has, respectively, the form:
\begin{equation*}
 S_{d}=\int \frac{\partial }{ \partial_{\mu}}U \frac{\partial }{ \partial ^{\mu}}U^* e^{ \gamma_R t}   d\vec{x}  dz.
 \end{equation*}
Part of the action, the variation of which gives a derivative for the field $\chi$ on the variable t:  
\begin{equation*}
S_{\chi}=\int (i (\frac{\partial }{ \partial t}\chi ) \chi^* - i (\frac{\partial  }{ \partial t}\chi^*) \chi )e^{ \gamma_R t}  d\vec{x}  dz %//
\end{equation*}
The term in action determining the phase interaction of the fields $\chi$ and U:
\begin{equation*}
S_{int}=\int ( I (\chi \chi^* )^2)e^{ \gamma_R t}  d\vec{x} dz.
\end{equation*}
Finally, the imaginary part of the action, which determines the exchange of quantity of matter:
\begin{equation*}
S_{f}=\int \alpha e^{ \gamma_R t}  d\vec{x}  dz.
\end{equation*}
and :
\begin{equation*}
S_{in}=\int \gamma_R(\partial_t U  U^*-\partial_t U^*  U)   e^{ \gamma_R t} d\vec{x}  dz.
\end{equation*}

The imaginary part of the action corresponds to the Hamilton function $H_{c}=\int \alpha d\vec{x} $.
Generally speaking, the introduction of an imaginary action is not a very good procedure \footnote{ this situation also arises when considering the usual linear absorption in a medium if one looks at the Schrodinger equation from the position of quantum mechanics, then the corresponding evolution operator is not Hermitian (and therefore he can not be associated with a physical quantity), and evolution is not unitary, however, in optics, the consideration of such equations is standard }, in that, for example, it can not automatically write out the integrals of motion, nevertheless, such models with imaginary action are discussed, for example, in work [\ref{Men}] or is used in describing scattering processes with absorption of so-called optical model of the nucleus. %[\ref{iden}]. 
For us, the immediate benefit of such a record is that it immediately shows how unambiguously the ionization coefficients are related to the plasma equations and the nonlinear absorption coefficient in the NSE equation for the light field.

Next, we turn to a number of special cases: 

a)  $S_{f}=0$,$S_{\chi}=0$,$S_{int}=0$,$\gamma_R=0$ - This situation is a medium with a cubic inertial nonlinearity in the absence of any dissipative mechanisms.
In this case, the original system is a real and Lagrangian system.

 Conservation laws corresponding to external global symmetries can be expressed by means of the energy-momentum tensor $P_{\nu \mu}$ constructed in the standard way in terms of the Lagrange function.

In particular, the law of conservation of energy (corresponding to the symmetry with respect to the shift in the evolutionary variable, which in this case is z), written in a differential form, will take the following form: 

\begin{equation*}
\frac{\partial h_{g}^{0}}{\partial z}=-\partial^j P_{z j}
\end{equation*}
	
Here

\begin{equation*}
 h_{g}^{0}\equiv P_{zz}=\partial_\mu U \partial^\mu U^*+ n_{ker}(I^2/{2}+(p_{\epsilon}^2-\Omega_R^2 \epsilon_{in}  ^2   )/2+\epsilon_{in} I(t) )
\end{equation*}

and
\begin{equation*}
p_{\epsilon}(t)=\int \theta(t-\zeta) exp(-\gamma_{R}(t-\zeta)) (-\Omega_{R}^2 \epsilon_{in}(\zeta)^2+I(\zeta) ) d \zeta
\end{equation*}

Accordingly, in the integral form we have:

\begin{equation*}
 H_{g}^{0} \equiv\int h_{g} d\vec{x}=H_{d}+H_{ker}+H_{in} =const
\end{equation*}

Next we will include in our consideration dissipative mechanisms associated with the oscillatory friction $ \gamma_R> 0 $. Then, after a series of transformations, one can obtain:
 
\begin{equation*}
\frac{ \partial H_{g}^{0}}{\partial z} =n_{ker}\int (\gamma_R p_{\epsilon}  (\int_{-\infty}^\infty \theta(x_0-\tau) R (x_0-\tau)s_{0}(\tau) d\tau)) d\vec{x}
\end{equation*}

Where   $ s_{\nu}=(\partial_{\nu} U  U^*-\partial_{\nu} U^*  U)/(2 i) $.  	This integral relation is a generalization of the widely known integral relation for a medium with instantaneous cubic nonlinearity [\ref{Tal}] (and as far as we know it is written out in explicit form for the first time), and can be similar to him (see for example [\ref{kand_num}]) be used to verify the correctness of numerical calculations, but already for a medium with inertial cubic nonlinearity.

b) We consider the case only with condition $S_{f}=0$. In this case, the action is real, and the interaction between the radiation and the field $\chi$ reduce to a fictitious phase rotation of the field $\chi$. This situation corresponds to the stage of nonstationary self-focusing in the medium by inertial cubic nonlinearity. 
Then it is not difficult to write down the integrals of motion. So the integrals of motion corresponding to the internal symmetry - the phase rotation, have the form:   

\begin{equation*}
    \chi \chi^*  =const
\end{equation*}
	
The law of conservation of the number of particles for plasma.

\begin{equation*}
\int (  U U^* ) d\vec{x} =const
\end{equation*}
	
The law of conservation of the number of particles for a light field.

This situation is of interest because in it for the Hamilton function  $H{_g}\equiv H_{d}+H_{ker}+H_{in}+H_{int}+H_{\chi}$   the simple relation: 
\begin{equation*}
 H{_g}\equiv\int h{_g}d\vec{x} =const
\end{equation*}	

And in this case the difference in the expression for $ H $ from the previous one is given by the term $H_{int}+H_{\chi}\equiv H_{int} \equiv -\int ( I (\chi \chi^* )^2) d\vec{x} $ , which in the case $ (\ chi \ chi ^ *) ^ 2 = const $ reduces to a linear combination of the integral of the motion $ H $ in an inertial cubic medium and the integration expressing the law of conservation of the number of particles. 

 c) Let us pass to the consideration of the general case. The integrands of motion for the number of particles in this case take the form: 

\begin{equation*}
 (\chi \chi^*  e^{\int_0^t{\alpha_{\rho} \rho d \xi }} )   =const
\end{equation*}

\begin{equation*}
\int ( U U^* e^{\int_0^z{\int\alpha_I I d \zeta d\vec{x}/(\int I  d\vec{x}) }}  ) d\vec{x} =const
\end{equation*}
For $H{_g}\equiv H_{d}+H_{ker}+H_{in}+H_{int}+H_{\chi}$  from the equations of dynamics in full form, one can obtain:\\

\begin{math}
\frac {\partial }{\partial z } H_g =\\-\int (i \alpha_{I} (U \Phi_{U}^*- U^* \Phi _{U})+i \alpha_{\chi}( \chi \frac {\delta G^*_{\chi }(I) }{\delta I } \frac {\partial s_{\nu}  }{\partial x_{\nu}  }- \chi^* \frac {\delta G_{\chi }(I) }{\delta I } \frac {\partial s_{\nu} }{\partial x_{\nu} })+\gamma_R p_{\epsilon} \frac {\partial G_{\epsilon }(I) }{\partial I } \frac {\partial s_{\nu} }{\partial x_{\nu} }) d\vec{x}
 \end{math}\\

Where $ \chi (I) $ is a well-known function of $ I $ in many cases, because for a number of important applications $ \psi_ {pl} ^ 1 = 0 $. In this case, the right-hand side has the form of an explicit function of the field $ U $.
After simple transformations it is possible to obtain:

\begin{equation*}
\frac {\partial H{_g}}{ \partial z} =-\int\alpha_I((2 h_d+I \frac{\delta h_{ker}}{\delta I})-\partial_{\nu} \partial^{\nu} I) -\chi \chi^* (\alpha_I+\bar{\alpha_{I \rho}}) d\vec{x}
\end{equation*}	

Where
\begin{equation*}
\bar{\alpha_{I \rho}}\equiv \partial_z\int_{0}^{t} (\alpha_{\rho})d\xi=-\int_{0}^{t} ((\alpha_{I}+\partial_{\nu} s^{\nu})\alpha_{I \rho}+(\chi \chi^*(\partial_z\int_{0}^{\xi} (\alpha_{\rho})d\zeta) )\alpha_{\rho \rho})d\xi
\end{equation*}	

This ratio is, again, useful for the possibility of checking and correcting numerical calculations already in the complete model of filamentation.\footnote{
It should be noted that in the literature, a more general equation is used to describe a broadband extremely short laser pulse than the nonlinear Schrodinger equation, namely a nonlinear equation of the generalized Kadomtsev-Petviashvili type [\ref{Skob},\ref{Boyd}]. As far as we know, its Lagrangian function is written out only for the case of local nonlinearity considered by us in the first section. The construction of the Lagrangian formulation, for this more general form of the equations, in the case of nonlocal effects, such as inertial cubic nonlinearity and plasma formation, requires a separate study that goes beyond the scope of this work.} 

\section*{Conclusion } 
In this paper, we propose a variational formulation of the system of equations for the nonlinear Schrodinger equation and the plasma density.
To proceed to the variational formulation, we needed to apply various methods, this is the method of integrating factors for solving the inverse variational problem and the method of auxiliary fields. So for the terms describing the effect of inertial cubic self-action, the transition to the variational formulation can be achieved by the method of integrating factors.  However, the compatibility condition for the equations for plasma and NSE can hardly be achieved by a similar method. The variational formulation in this case can be achieved by introducing an auxiliary field % with a purely gauge mode, namely, a transition to an equivalent equation, but written for the complex amplitude of the localization density of the plasma, which means embedding the original real dynamical system in a complex dynamical system.
Also in this paper we analyze equations for NLES integrals in various of its formulations including a complete nonstationary model with nonlinear dissipation . 
Knowledge of these conservation laws is a valuable tool not only for checking the correctness of numerical schemes widely used in numerical modeling of NLES, but also for developing fully conservative methods of numerical integration of these equations.

%\appendix

%\section{The First Appendix}

%The appendix fragment is used only once. Subsequent appendices can be created

%using the Section Section/Body Tag.
\end{document}